# Ударная электромагнитная волна большой амплитуды в нелинейной линии передачи на основе распределенного полупроводникового диода.

## А. С. Кюрегян

Всероссийский Электротехнический институт им. В. И. Ленина, 111250, Москва, Россия

E-mail: semlab@yandex.ru

Получено аналитическое решение задачи о возникновении и распространении ударной электромагнитной волны большой амплитуды в полосковой линии передачи (ЛП) на основе распределенного полупроводникового диода, которое корректно учитывает ее нелинейность, диссипацию и временную дисперсию. Результаты теории использованы для оценки параметров ЛП как обострителя фронта импульса напряжения, подаваемого на вход линии.

**1. Введение**

Давно известно, что в нелинейных линиях передачи (ЛП) при определенных условиях возникают ударные электромагнитные волны, которые используются, в частности, для формирования импульсов напряжения с крутыми фронтами [1-6]. Нелинейными элементами могут быть распределенные вдоль линии индуктивность $L$, зависящая от тока $I$ при использовании ферритов, и/или емкость $C$, зависящая от напряжения $U$ при использовании ферроэлектриков, полупроводниковых МДП-структур, диодов с барьером Шоттки или с $p^+$-$n$-переходом. Именно этот последний вариант ЛП теоретически исследован в настоящей работе применительно к проблеме обострения импульсов напряжения большой амплитуды.

Такую ЛП можно представить в виде множества последовательно соединенных ячеек, эквивалентная схема которых изображена на Рис. 1a. Распространение по ней электромагнитной волны вдоль оси $z$ описывается при известных [7] условиях системой телеграфных уравнений

$$\frac{\partial U}{\partial z} + L\frac{\partial I}{\partial t} + R_L I = 0, \qquad \frac{\partial I}{\partial z} + \frac{\partial Q}{\partial t} + GU = 0 \qquad (1)$$

где $I = I(t,z)$ - ток вдоль линии, $L$ и $R_L$ - постоянные погонные индуктивность и сопротивление линии, $G^{-1}$ - сопротивление утечки. Нелинейность соотношения между напряжением $U = U(t,z)$ и погонной плотностью заряда $Q = Q(t,z)$ линии является причиной возникновения ударной волны. Среди использованных ранее соотношений наиболее общее и подходящее для нашего случая было предложено авторами работы [8]. Его можно представить в виде двух равенств

$$U = Q/C_w(Q) + U_n, \qquad U_n = R_n \partial(Q - C_n U_n)/\partial t, \qquad (2)$$

где $C_w$ - барьерная емкость диода, $U_n$ - падение напряжения на емкости $C_n$ и сопротивлении $R_n$ нейтральной области $n$-слоя диода (см. Рис. 1b). Авторы остальных известных нам работ [1,2,9-13] либо использовали неподходящие для нашего случая зависимости $C_w(Q)$, либо пренебрегали величинами $C_n$ и/или $R_n$, а те, кто их учитывал (в том числе и авторы работы [8]), полагали $C_n$ и $R_n$ постоянными. На самом деле при большой амплитуде волны величины $R_n$ и $C_n$ явно зависят от времени из-за уменьшения толщины $(d-w)$ нейтральной области $n$-слоя, а $R_n$ зависит еще и от плотности тока $J = \partial Q/\partial t$, протекающего поперек ЛП, вследствие уменьшения подвижности электронов с ростом напряженности электрического поля.

Способ учета этих особенностей, позволяющий корректно описать нелинейность, диссипацию и временную дисперсию активного *n*-слоя диода, изложен в разделе 2 настоящей статьи. В разделе 3 дано аналитическое описание реальной структуры фронта стационарной ударной волны в ЛП на основе распределенного диода, в разделе 4 – описаны процессы формирования и затухания ударной волны при подаче на вход ЛП относительно медленно нарастающего напряжения, а в Заключении дана оценка параметров ЛП как обострителя фронта импульса напряжения.

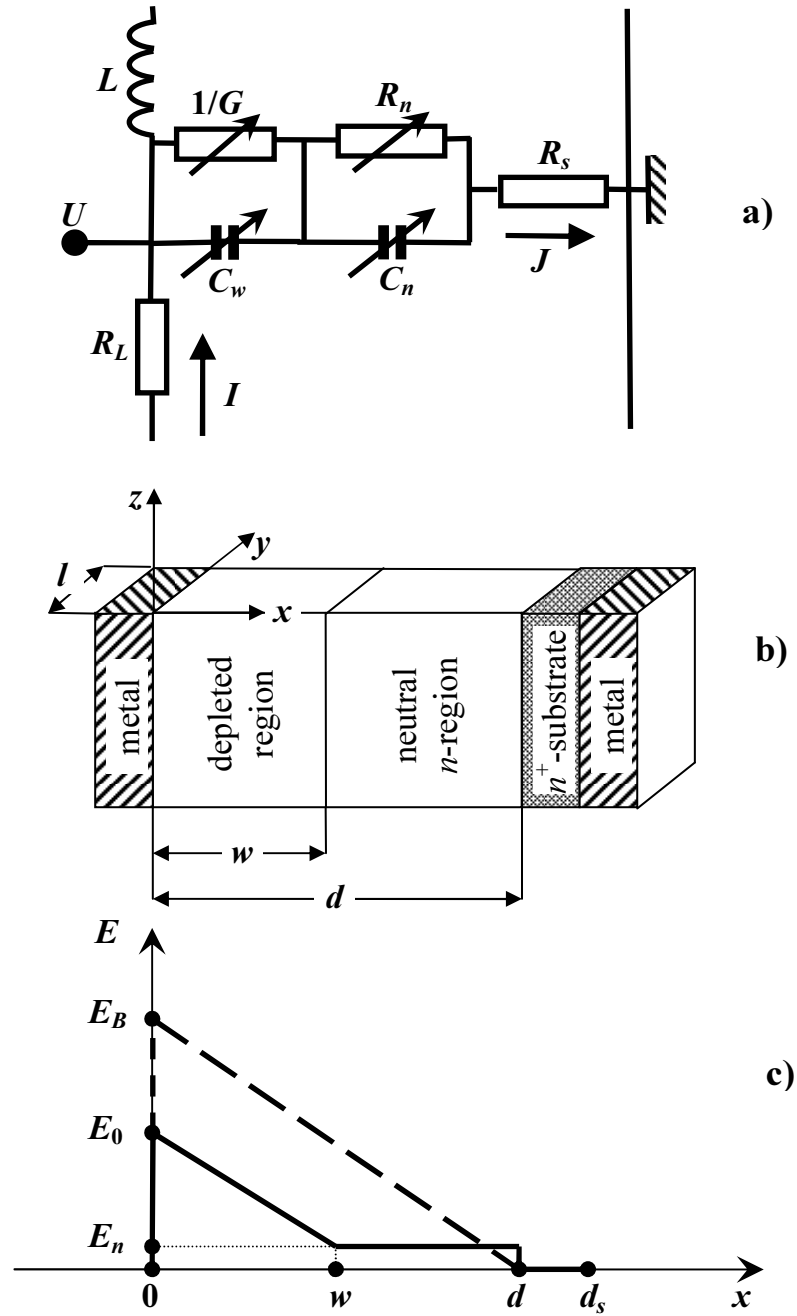

Рис. 1. Эквивалентная схема ячейки нелинейной линии передачи (a), поперечное сечение полоскового диода (b) и распределение электрического поля в $p^+$-$n$-$n^+$-структуре (c). $C_w$ и $G$ - емкость и проводимость истощенной области полупроводника $(0 < x < w)$, $C_n$ и $R_n$ - емкость и сопротивление нейтральной области *n*-слоя $(w < x < d)$, $R_s$ - сопротивление подложки $(d < x < d_s)$. Тонкий $p^+$-слой не показан.

## 2. Нелинейность и временная дисперсия диодов.

Рассмотрим полосковую линию передачи, состоящую из протяженного вдоль оси $z$ обратно смещенного $p^+$-$n$-$n^+$-диода, поперечное сечение которого схематически изображено на Рис. 1b. Будем считать, что

- $n^+$-подложка толщиной $d_s$ однородно легирована донорами с концентрацией $N_s$,
- активная $n$-область толщиной $d$ однородно легирована донорами с концентрацией $N \ll N_s$,
- $p^+$-слой (на рисунке не показан) толщиной $d_p \ll d$ легирован акцепторами с концентрацией $N_p \gg N$,
- ширина полоски $l$ и толщина фронта $\delta_f$ ударной волны, распространяющейся по линии, много больше $(d+d_s)$,
- проводимость $G$ истощенной области полупроводника $(0 < x < w)$ пренебрежимо мала.

Тогда поперечное электрическое поле $E$ в $n$-области практически не зависит от $y$, а вдоль оси $x$ распределено так, как изображено на Рис. 1c. Разность потенциалов между электродами $U(t,z)$, включающая и контактную разность потенциалов, связана с напряженностями поля $E_0(t,z)$, $E_n(t,z)$, $E_s(t,z)$ в плоскостях $x=0$, $x=d$, $x=d+d_s$ и с толщиной истощенной области $w(t,z)$ очевидными соотношениями

$$U = (E_0 - E_n)\frac{w}{2} + E_n d + E_s d_s, \tag{3}$$

$$E_0 = \frac{e}{\varepsilon} N w + E_n, \tag{4}$$

где $\varepsilon$ - диэлектрическая проницаемость полупроводника, $e$ - элементарный заряд. Далее для определенности мы будем считать, что в статическом случае (то есть при $E_n = E_s = 0$) $w = d$ и $E_0 = E_B = eNd/\varepsilon$ при напряжении лавинного пробоя $U_B$. Электрические поля $E_n$, $E_s$ в областях $w < x < d$, $d < x < d+d_s$ не зависят от $x$ и обусловлены протекающим через них током с линейной плотностью

$$J = \frac{\partial Q}{\partial t} = l\varepsilon \frac{\partial E_0}{\partial t} = l\left[eNv(E_n) + \varepsilon\frac{\partial E_n}{\partial t}\right] = leN_s \mu_{es} E_s, \tag{5}$$

где $Q(t,z) = l\varepsilon E_0(t,z)$ - погонная плотность заряда в плоскости $x=0$, $v(E)$ - дрейфовая скорость электронов, для которой мы будем использовать аппроксимацию

$$v(E) = v_e \frac{1+(E/E_1)^2}{E_e/E + 1 + (E/E_2)^2}, \tag{6}$$

$v_e = \mu_e E_e$, $\mu_e$ и $\mu_{es}$ - низкополевые подвижности электронов в $n$-области и подложке. Она хорошо описывает реальные зависимости $v(E)$ в Si [14] (где скорость практически постоянна в сильном поле) и 4H-SiC [15] (где скорость уменьшается с ростом $E$ при $E > 200$ кВ/см) при значениях подгоночных параметров, приведенных в таблице 1.

Таблица 1.

| Параметр | $v_e$, см/мкс | $E_e$, кВ/см | $E_1$, кВ/см | $E_2$, кВ/см | $E_i$, МВ/см | $U_i$, В |
|----------|---------------|--------------|--------------|--------------|--------------|----------|
| Si       | 10,0          | 7,1          | $\infty$     | $\infty$     | 1,43         | 1,83     |
| 4H-SiC   | 18,5          | 21           | 900          | 550          | 15,9         | 2,15     |

Первое слагаемое в квадратных скобках формулы (5) соответствует току проводимости через сопротивление $R_n$, а второе – току смещения через емкость $C_n$. Так как обычно $N_s \gg N$, то в подложке можно пренебречь током смещения и снижением подвижности в поле $E_s \ll E_e$. Из (4) и (5) следует, что

$$\partial w/\partial t = v(E_n), \qquad (7)$$

то есть граница $x = w$ между истощенной и нейтральной областями $n$-слоя движется со скоростью $v(E_n)$, изменяя тем самым эффективные значения $R_n$ и $C_{w,n}$. Формулы (3)-(5) и (7) задают в неявном виде соотношение между $U$ и $Q$ и являются аналогом равенств (2), однако только они корректно и исчерпывающе описывают особенности нелинейности, диссипации и временной дисперсии активного $n$-слоя диода.

### 3. Стационарная ударная волна.

Стационарная ударная волна, строго говоря, может существовать только в бесконечно длинной ЛП с пренебрежимо малыми сопротивлением $R_L$ и проводимостью $G$. В этом случае в системе координат $\tilde{z} = z - u_f t$, движущейся вместе с фронтом с постоянной скоростью $u_f$, уравнения (1) принимают вид

$$\frac{\partial U}{\partial \tilde{z}} - Lu_f \frac{\partial I}{\partial \tilde{z}} = 0, \qquad \frac{\partial I}{\partial \tilde{z}} - u_f \frac{\partial Q}{\partial \tilde{z}} = 0 \qquad (8)$$

Интегрирование (8), с граничными условиями

$$U(\pm\infty) = U_\pm, \quad Q(\pm\infty) = Q_\pm, \quad I(+\infty) = 0 \qquad (9)$$

приводит к соотношениям

$$Q(\tilde{z}) - Q_+ = \overline{C}\left[U(\tilde{z}) - U_+\right], \qquad U(\tilde{z}) - U_+ = Lu_f I(\tilde{z}), \qquad (10)$$

где эффективная погонная емкость

$$\overline{C} = \frac{Q_- - Q_+}{U_- - U_+} = \frac{\varepsilon l}{\overline{w}}, \qquad (11)$$

$\overline{w} = (w_+ + w_-)/2$, $w_\pm = 2U_\pm/E_\pm$, $E_\pm = E_B\sqrt{U_\pm/U_B} = Q_\pm/l\varepsilon$.

Далее мы будем считать, что параметры подложки $N_s, d_s$ таковы, что выполняются два сильных неравенства

$$E_s d_s \ll E_n d \quad \text{и} \quad d_s \ll h_s = 2\sqrt{t_f/\mu_0 \sigma_s}, \qquad (12)$$

где $h_s$ - глубина проникновения магнитного поля в подложку с проводимостью $\sigma_s = e\mu_{es}N_s$ за время $t_f = \delta_f/u_f$ нарастания тока на фронте волны, $\mu_0$ магнитная постоянная. В этом случае диссипация в подложке пренебрежимо мала, погонная индуктивность линии $L = \mu_0(d + d_s)/l$ и скорость фронта

$$u_f = (L\overline{C})^{-1/2} = c_s\sqrt{\frac{\overline{w}}{d + d_s}} \qquad (13)$$

может быть значительно меньше скорости света $c_s = 1/\sqrt{\varepsilon\mu_0}$ в полупроводнике [16-19]. Волновое сопротивление такой линии равно

$$Z = \frac{U(\tilde{z}) - U_+}{I(\tilde{z})} = \sqrt{\frac{L}{\overline{C}}} = Z_s \frac{\sqrt{\overline{w}(d + d_s)}}{l} \qquad (14)$$

где $Z_s = \sqrt{\mu_0/\varepsilon} \approx 110$ Ом в кремнии.

Используя (3) без малого слагаемого $E_s d_s$ вместе с первым из равенств (10) легко показать, что при этих условиях

$$E_n(E_0) = E_0 - E_B + \sqrt{(E_B - E_0)^2 + (E_- - E_0)(E_0 - E_+)}, \qquad (15)$$

$$U(E_0) = U_B E_B^{-2} \left[ E_0(E_+ + E_-) - E_+ E_- \right]. \qquad (16)$$

Как и следовало ожидать, при $E_0 \to E_\pm$ (то есть вдали от фронта, где $J \to 0$) $U_0 \to U_\pm$, а напряженность поля $E_n \to 0$ и достигает максимума

$$\hat{E}_n = \frac{\overline{E}^2}{2(E_B - \overline{E})} \qquad (17)$$

на фронте при

$$E_0 = \hat{E}_0 = \frac{E_B E_+ + E_B E_- - E_+ E_- - \overline{E}^2}{2(E_B - \overline{E})}, \qquad (18)$$

где $\overline{E} = (E_+ + E_-)/2$. Зависимости $E_n(E_0)$ приведены на Рис. 2. Видно, что при больших $U_-$ и малых $U_+$ максимальная напряженность поля в нейтральной области $n$-слоя значительно превосходит величину $E_e \leq 0.1 E_B$, при которой зависимость $v(E)$ начинает отличаться от линейной. Из формул (4) и (15) можно получить максимальное значение $\hat{U}_n$ падение напряжения на нейтральной части $n$-слоя $U_n = E_n(d - w)$. Точная формула очень громоздкая, поэтому мы приводим здесь приближенную формулу $\hat{U}_n = 4U_-/27$ для случая $U_+ = 0$, которая обеспечивает погрешность менее 15% и дает точное значение $\hat{U}_n$ при $U_- = U_B$.

В движущейся системе координат уравнение (5) принимает вид

$$u_f \frac{\partial(E_0 - E_n)}{\partial E_0} \frac{\partial E_0}{\partial \tilde{z}} + \frac{E_B}{d} v[E_n(E_0)] = 0 \qquad (19)$$

и интегрируется с учетом (15) при любой зависимости $v(E)$:

$$\tilde{z} = u_f d \int_{E_0}^{\overline{E}} \frac{(1 - \overline{E}/E_B) dE}{v[E_n(E)] \sqrt{(E - E_B)^2 + (E_- - E)(E - E_+)}}, \qquad (20)$$

начало координат на оси $\tilde{z}$ выбрано в точке, где $E_0 = \overline{E}$. Примеры распределений напряжения $U$ вдоль линии, рассчитанные по формулам (16),(20), приведены на Рис. 3. Как видно, структуры фронта в ЛП на основе Si и 4H-SiC практически совпадают, несмотря на существенное различие между типами зависимостей $v(E)$ в этих материалах. Если $E_- < E_B$, то интеграл в (20) расходится логарифмически при $E_0 \to E_\pm$, так как при этом $v(E_n) \propto |E_0 - E_\pm|$.

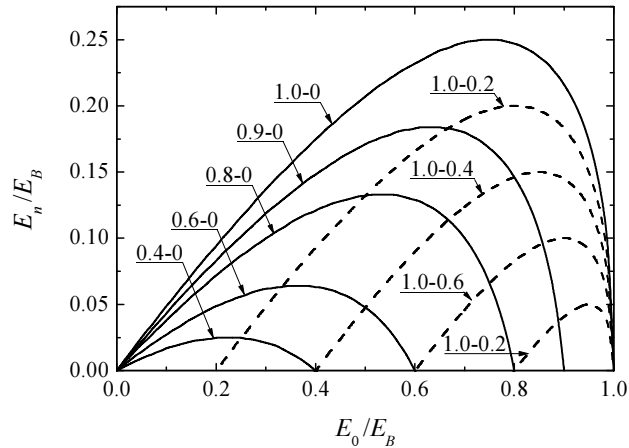

Рис. 2. Графики функции $E_n(E_0)$, рассчитанные по формуле (15). Числа у кривых указывают значения отношений $E_-/E_B$ (первое) и $E_+/E_B$ (второе).

Поэтому при удалении от фронта $E_0$ приближается к $E_{\pm}$ по закону

$$E_0(\tilde{z}) - E_{\pm} \propto \pm \exp(\mp \tilde{z}/\lambda), \qquad (21)$$

где $\lambda = 2u_f t_M \dfrac{E_B - \overline{E}}{E_- - E_+}$, $t_M = \varepsilon/eN\mu_e$ - время максвелловский релаксации в $n$-области. Вырожденный случай $E_- = E_B$ отличается тем, что в формуле (21) нужно заменить $\lambda$ на $\lambda/2$ при $E_0 \to E_-$. Из (20) следует, что характерное время нарастания напряженности поля на фронте

$$t_f = \frac{E_- - E_+}{u_f}\frac{d\tilde{z}}{dE_0}\bigg|_{\tilde{z}=0} = \frac{d(E_- - E_+)(E_B - \overline{E})}{E_B v[E_n(\overline{E})]}\sqrt{\frac{2}{(E_B - E_+)^2 + (E_B - E_-)^2}}. \qquad (22)$$

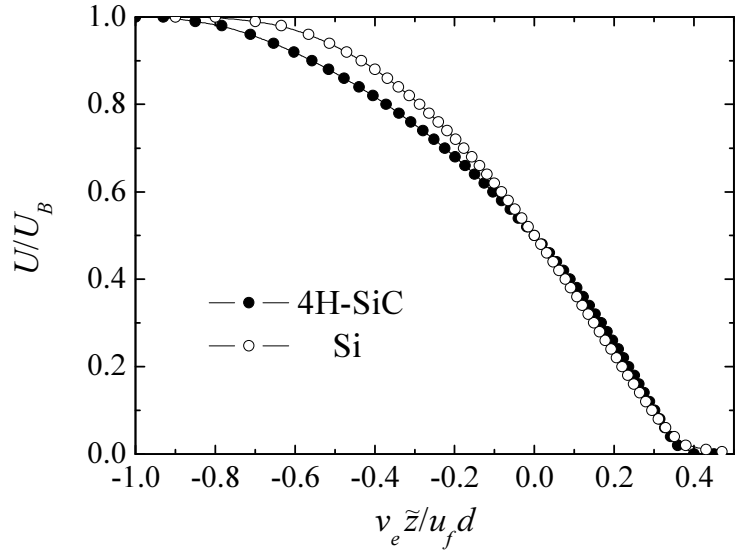

Рис. 3. Распределения напряжения $U$ на фронте ударной волны в кремниевой (светлые символы) и карбид-кремниевой (темные символы) линиях, рассчитанные по формулам (16) и (20) при $U_- = U_B$ и $U_+ = 0$.

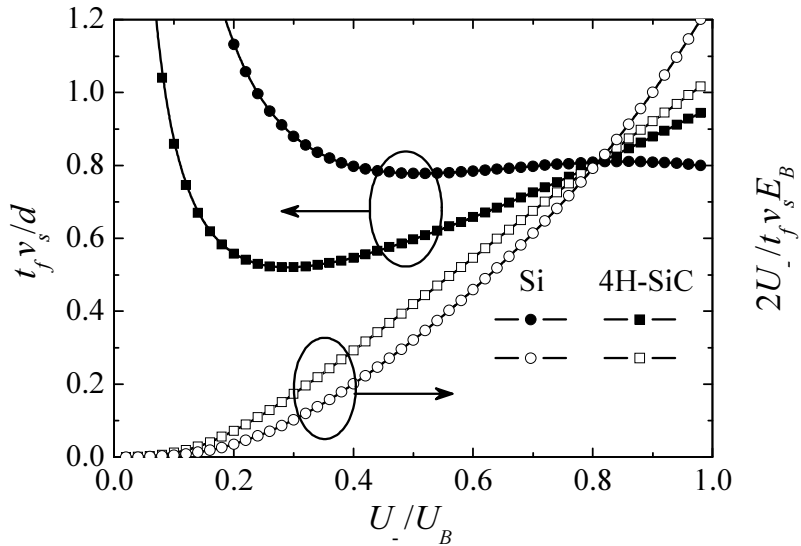

Рис. 4. Зависимости нормированных длительности $t_f$ фронта ударной волны (темные символы) и скорости нарастания напряжения $U_-/t_f$ на фронте (светлые символы) от $U_-$, рассчитанные по формуле (22) для кремниевой и карбид-кремниевой линий при $U_+ = 0$.

Примеры зависимостей $t_f$ от $U_-$, рассчитанные по этой формуле, приведены на Рис. 4. Постоянство $t_f$ в Si и увеличение $t_f$ в 4H-SiC при больших $U_-$ обусловлено соответственно насыщением зависимости $v(E)$ и уменьшением $v(E)$ в сильных полях $E_n$. Однако характерные скорости нарастания напряжения $U_-/t_f$ на фронте в обоих случаях монотонно увеличиваются с ростом $U_-$ и максимальны при $U_- = U_B$.

При прохождении фронта ударной волны в нейтральной части $n$-слоя диода рассеивается мощность с линейной плотностью $P_n = lE_n(d-w)eNv(E_n)$. Используя (4) и (5), эту формулу можно представить в виде

$$P_n = ld\varepsilon E_n \left(1 - \frac{E_0 - E_n}{E_B}\right)\left(1 - \frac{\partial E_n}{\partial E_0}\right)\frac{\partial E_0}{\partial t}. \tag{23}$$

Интегрируя (23) по времени с учетом (15) нетрудно получить линейную плотность энергии, рассеянной на фронте после прохождения волны:

$$\omega_n = \int_{-\infty}^{\infty} P dt = ld\frac{\varepsilon}{E_B}\left[\frac{(E_B - E_+)^3 - (E_B - E_-)^3}{3} - (E_- - E_+)(E_B - \bar{E})^2\right]. \tag{24}$$

При $E_+ = 0$ формула (24) упрощается и принимает следующий вид:

$$\omega_n = ld\varepsilon \frac{E_-^3}{12 E_B} \tag{25}$$

Как видно, в этом случае $\omega_n$ в $2\sqrt{U_B/U_-}$ раза меньше линейной плотности энергии электрического поля $\omega_- = ld\varepsilon E_-^2/6$ в ЛП, заряженной до напряжения $U_-$.

Следует отметить неожиданное свойство изучаемой ЛП: величина подвижности электронов и характер ее зависимости от напряженности поля определяют толщину фронта $\delta_f$ и время нарастания напряжения $t_f$ (см. (20)-(22)), но никак не влияют на максимальное падение напряжения $\hat{U}_n$ и линейную плотность энергии потерь $\omega_n$ в нейтральной части $n$-слоя.

Наибольший практический интерес представляет случай $E_+ = 0$, $E_- = E_B$, когда[1] $t_f \approx \xi d/v_e$ и толщина фронта $\delta_f = u_f t_f \approx \xi c_s d/v_e \sqrt{2\gamma}$, где $\gamma = (1 + d_s/d)$. Используя полученные результаты можно показать, что в этом случае оба условия (12) выполняются при

$$d_s/d << \min\left(\frac{N_s\mu_{es}}{N\mu_e}, \frac{2c_s}{v_e}\sqrt{\xi\frac{N\mu_e E_e}{N_s\mu_{es} E_B}}\right). \tag{26}$$

Правая часть этого неравенства максимальна при

$$\frac{N_s\mu_{es}}{N\mu_e} = \left[\xi\frac{E_e}{E_B}\left(2\frac{c_s}{v_e}\right)^2\right]^{1/3} \equiv \gamma_0, \tag{27}$$

откуда следует соотношение

$$\frac{\delta_f}{d + d_s} \approx \left(\frac{\gamma_0}{\gamma}\right)^{3/2}\sqrt{\frac{E_B}{8E_e}}.$$

Так как обычно $E_e < 0.1 E_B$, то при условии (26) или, что то же самое $\gamma << \gamma_0$, выполняется сильное неравенство $\delta_f >> (d + d_s)$, которое оправдывает применение телеграфных уравнений (1) и одномерных соотношений (3)-(7) для анализа структуры фронта ударной волны.

---

[1] Множитель $\xi$ близок к единице; он равен 0,8 для Si и 0,96 для $H-SiC (см. Рис. 4).

## 4. Формирование и затухание ударной волны.

Точный учет временной дисперсии и диссипации с помощью формул (3)-(5) и (7) необходим только вблизи фронта, где плотность тока $J$ очень велика и быстро изменяется со временем. При удалении от фронта (при $|\tilde{z}| > \lambda$) плотность тока $J$ и скорость его изменения экспоненциально уменьшаются. Поэтому в формулах (3),(4) можно положить $E_n = E_s = 0$ и использовать вместо них одно простое соотношение

$$\frac{U}{U_B} \equiv V = \left(\frac{Q}{Q_B}\right)^2, \qquad (28)$$

которое правильно описывает нелинейность емкости диода. Разумеется, при этом не учитывается дисперсия и диссипация на фронте. Кроме того, для упрощения задачи, будем считать погонное сопротивление $R_L$ постоянной величиной, пренебрегая тем самым скин-эффектом в металле электродов. Тогда из (1) следует уравнение

$$\frac{\partial^2 q}{\partial \theta^2} + (r+g)\frac{\partial q}{\partial \theta} + rgq = \frac{\partial}{\partial \chi}\left(q\frac{\partial q}{\partial \chi}\right), \qquad (29)$$

которое можно использовать для качественного анализа процессов формирование и затухание ударной волны в ЛП, на вход которой подается медленно нарастающее напряжение. В (29) использованы безразмерные переменные

$$q = Q/Q_B, \quad \theta = t/t_0, \quad \chi = \frac{z}{t_0}\sqrt{\frac{L\overline{C}_B}{2}},$$

параметры $r = t_0 R_L / L$, $g = t_0 / \tau_g$, характеризующие диссипацию за фронтом и учтено, что в обратно смещенных диодах линейная плотность тока утечки $GU = eNlw/\tau_g = Q/\tau_g$, где $\tau_g = t_g N/n_i$, $t_g$ - генерационное время жизни, $n_i$ - собственная концентрация в полупроводнике, $\overline{C}_B = 2\varepsilon l/d$. Далее мы будем использовать простейшие начальное и граничное условия

$$q(0,\chi) = 0 \quad \text{и} \quad q(\theta,0) = \begin{cases} \sqrt{\theta} & \text{при} \quad \theta < 1 \\ 1 & \text{при} \quad \theta > 1 \end{cases}, \qquad (30)$$

соответствующие линейному нарастанию напряжения на входе в незаряженную ЛП от нуля до $U_B$ за время $t_0$. Обычно параметр $rg/(r+g)^2$, характеризующий относительный вклад в затухание волны двух различных механизмов, очень мал вследствие неравенства $L/R_L \ll \tau_g \sim 1$ мс, поэтому можно считать $g = 0$.

Если еще и сопротивление $R_L$ пренебрежимо мало, то нетрудно [3-6,20] найти частное решение задачи Коши (29),(30), которое описывает распространение так называемой простой волны [20], возбуждаемой внешним источником напряжения на входе (то есть при $z = 0$) в незаряженную ЛП:

$$\chi = \sqrt{q}\left(\theta - q^2\right). \qquad (31)$$

Эта формула применима при $\chi_\theta \leq \chi \leq \chi_f(\theta)$ и $1 \geq q \geq q_f(\theta)$, где $\chi_\theta = \theta H(\theta-1)$, $H(x)$ - ступенчатая функция Хевисайда, $\chi_f$ и $q_f$ положение и амплитуда разрыва, который в этом приближении образуется вместо фронта конечной толщины и устраняет нефизичную двузначность зависимости (31) $q$ от $\chi$ [20]. Если $t_f \ll t_0$, то сам фронт является квазистационарным, так что для вычисления его структуры и скорости можно использовать соотношения, полученные в предыдущем разделе, полагая $U_- = U_B q_f^2$. В частности, безразмерная скорость разрыва $v_f = \partial \chi_f / \partial \theta$ должна быть равна $\sqrt{q_f/2}$, поэтому дифференцирование (31) по $\theta$ при $q = q_f$ приводит к уравнению

$$\left(\theta - 5q_f^2\right)\frac{\partial q_f}{\partial \theta} = \left(2-\sqrt{2}\right)q_f,$$

решая которое с начальным условием $q_f(0) = 0$, получим

$$q_f = \sqrt{\frac{\theta}{\chi_1+1}}, \qquad \chi_f = \chi_1\left(\frac{\theta}{\chi_1+1}\right)^{5/4}, \qquad (32)$$

$$q_f = \left(\chi_f/\chi_1\right)^{2/5}, \qquad (33)$$

где $\chi_1 = 2\sqrt{2}/\left(5-2\sqrt{2}\right) \approx 1.3$. Формулы (31)-(33) отлично описывает численное решение задачи Коши (29),(30) без учета затухания, приведенное на Рис. 5.

При $\theta > 1$ появляется область $0 < \chi < \chi_\theta$, в которой $q = 1$, тогда как при $\chi_\theta < \chi < \chi_f$ профиль волны по-прежнему описывается формулой (31). Правая граница $\chi_\theta$ движется со скоростью $\partial \chi/\partial \vartheta|_{q=1} = 1$ и догоняет фронт в плоскости $\chi = \chi_1$ в момент $\theta = \theta_1 = 1+\chi_1$. После этого возникает стационарная волна с амплитудой $q_f = 1$ и безразмерной скоростью фронта $v_f = \sqrt{1/2}$. Используя (25) и (33) нетрудно показать, что за время $\theta_1$ в нейтральной части $n$-слоя на фронте рассеивается энергия

$$\Omega_1 \equiv \int_0^{z_1} \omega_n\left[q_f(z)\right]dz = \frac{5}{66}lz_1\varepsilon E_B U_B, \qquad (34)$$

которая в 4,4 раза меньше энергии электрического поля $\omega_{-z_1}$, запасенной к этому времени в линии.

При конечном сопротивлении $R_L$ аналитическое решение задачи Коши (29),(30) неизвестно, даже если считать $R_L = const$. Результаты численного решения вполне ожидаемы: омические потери в электродах приводят к затуханию волны, как это изображено на Рис. 6. Амплитуда скачка напряжения достигает максимума $V_m = q_m^2$ при $\chi = \chi_m$ в момент $\theta = \theta_m$. Поэтому оптимальная длина ЛП, используемой для обострения фронта входного импульса, равна $z_m = \chi_m t_0 \sqrt{2/L\overline{C}_B}$.

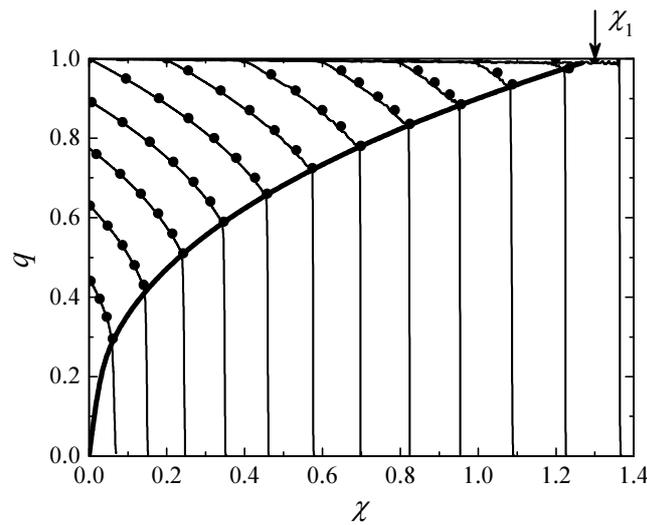

Рис. 5. Профили незатухающей ударной волны в моменты $\theta = 0.2; 0.4....2.4$ при $r = g = 0$. Тонкие сплошные линии - численное решение задачи Коши (29),(30), темные символы - расчет $q(\theta, \chi)$ по формуле (31), толстая линия - зависимость $q_f(\chi_f)$, рассчитанная по формуле (33).

Величины $\chi_m$ и $\theta_m$ очень слабо уменьшаются с ростом $r$ (см. Рис. 7) и с достаточной для практических целей точностью могут считаться постоянными и равными $\chi_1$ и $\theta_1$ соответственно. Максимальная амплитуда разрыва $V_m = U_m/U_B$, естественно, тоже уменьшается с ростом $r$, но медленнее, чем нарастает омическое падение напряжения $U_r$ на электродах между входом в линию и плоскостью $\chi = \chi_m$, равное в рассматриваемом здесь случае [21]

$$\frac{U_r(t_m)}{U_B} \equiv V_r(\theta_m) = 1 - \frac{1}{r}(e^r - 1)e^{-r\theta_m}\bigg|_{t_m \ll t_r} \approx 4.6\frac{t_0}{t_r}. \tag{35}$$

где $t_r = \sigma_M \mu_0 d_M (d + d_s)$, $\sigma_M$ и $d_M$ - проводимость и толщина электродов соответственно. Причина этого состоит в том, что затухание приводит к уменьшение со временем тока $I$ вдоль линии, так что второе слагаемое в (1) частично компенсирует «омическое» поле $R_L I$ и замедляет уменьшение напряжения $U$ с ростом $z$.

Сопротивление $R_L$ уменьшается с ростом $d_M$ и $\sigma_M$, но ограничено снизу скин-эффектом, учет которого приводит к тому, что вместо (29) получается нелинейное интегро-дифференциальное уравнение. Его решение даже численными методами представляет собой нетривиальную задачу. Однако при сильном скин-эффекте (то есть при $d_M > 2\sqrt{t_m/\sigma_M \mu_0}$) для омического падения напряжения $U_\sigma$ вдоль электродов между входом в линию и фронтом была получена точная формула [21]. При малом затухании в электродах, когда $t_0 \ll t_\sigma = \pi \sigma \mu_0 (d + d_s)^2$ ее можно представить в виде

$$\frac{U_\sigma(t)}{U_B} \equiv V_\sigma(\theta) = \frac{4}{3}\sqrt{\frac{t_0}{t_\sigma}}\left[\theta^{3/2} - (\theta - 1)^{3/2} H(\theta - 1)\right]_{t=t_m} \approx 2.67\sqrt{\frac{t_0}{t_\sigma}}. \tag{36}$$

Очевидно, при $V_\sigma(\theta_m) \ll 1$ скин-эффект не может привести к изменению соотношений $\chi_m \approx \chi_1$ и $\theta_m \approx 1 + \chi_1$, использование которых приводит к последнему приближенному равенству в (36).

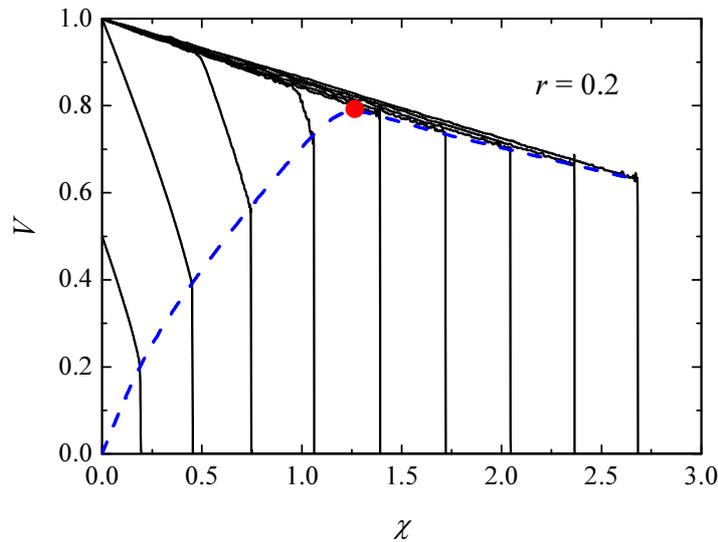

Рис. 6. Сплошные линии - профили затухающей ударной волны в моменты $\theta = 0.5; 1.0 .... 4.5$, полученные путем численного решения задачи Коши (29),(30), при $r = 0.2$ и $g = 0$. Штриховая линия – зависимость амплитуды разрыва от его положения. Точкой отмечено положение $\chi_m$ и амплитуда $V_m = q_m^2$ максимального разрыва в момент $\theta = \theta_m = 2.3$.

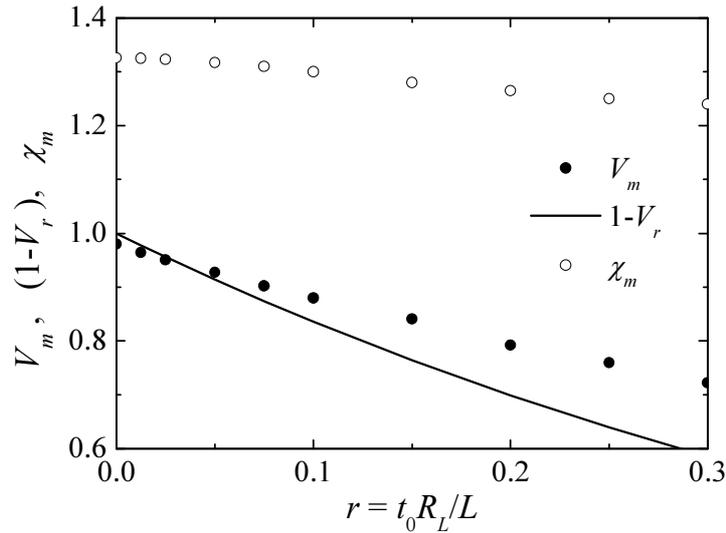

Рис. 7. Зависимости положения $\chi_m$ и амплитуды $V_m = q_m^2$ максимального разрыва от безразмерного сопротивления $r$. Символы – результаты численного решения, сплошная линия – расчет $(1-V_r)$ по формуле (35) при $\theta_m = 2.3$.

## 5. Заключение.

Полученные выше результаты позволяют оценить достижимые параметры ЛП как обострителя фронта импульса напряжения. В качестве примера проделаем это для ЛП на основе Si и 4H-SiC, с помощью которых нужно сформировать импульс с фронтом $t_f = 0.2$ нс и коэффициентом обострения $k = t_0/t_f = 10$. Эти условия позволяют найти все необходимые параметры линии:

- толщину n-слоя $d = t_f v_e/\xi$;
- напряжение пробоя $U_B$ и напряженность поля $E_B$, связанные с $d$ соотношениями $U_B = E_B d/2$ и $E_B = E_i/\ln(U_B/U_i)$, где $E_i$ и $U_i$ - подгоночные параметры, приведенные Таблице 1 для Si [22] и 4H-SiC [23];
- величину $\gamma_0$ по формуле (27) и полную толщину полупроводника $\gamma d$;
- оптимальную длину линии $z_m = \chi_m t_0 c_s/\sqrt{\gamma}$;
- минимальную толщину $d_M = 2\sqrt{\theta_m t_0/\sigma_M \mu_0}$ электродов, при которой их сопротивление определяется скин-эффектом;
- нормированное падение напряжения вдоль электродов $V_m = U_\sigma(\theta_m t_0)/U_B$ по формуле (36);
- нормированную ширину линии $l/\gamma d = Z_z/Z\sqrt{2\gamma}$.

Таблица 2.

| Параметр | $E_B$, МВ/см | $U_B$, кВ | $d$, мкм | $\gamma d$, мкм | $z_m$, см | $d_M$, мкм | $V_m$ | $l/\gamma d$ |
|---|---|---|---|---|---|---|---|---|
| Si | 0,27 | 0,37 | 27 | 290 | 6,9 | 22 | 0,037 | 4,7 |
| 4H-SiC | 2,1 | 4,48 | 43 | 240 | 10,6 | 22 | 0,046 | 7,2 |

Результаты расчетов при $\gamma = \gamma_0/4$, $\sigma_M = 3\cdot 10^5$ См/см и $Z = 5$ Ом, приведенные в Таблице 2, позволяют сделать следующие выводы. Во-первых, необходимая длительность фронта $t_f$ однозначно определяет его максимальную амплитуду $U = U_B$. По этому показателю карбид кремниевые ЛП обладают явным преимущество над кремниевыми. Причиной этого, очевидно, является гораздо более высокая электрическая прочность карбида кремния. Во-вторых, в рассмотренном случае омическое падение напряжения вдоль электродов $U_\sigma$ много меньше $U_B$ и даже почти в три раза меньше $\hat{U}_n$. Поэтому общие потери энергии в ЛП за время $t_m$ лишь немного превосходят $\Omega_1$. В-третьих, увеличение коэффициента обострения $k$ при заданной длительности фронта приводит к росту минимальной толщины $d_M$ электродов, напряжения $U_\sigma$ и, особенно быстрому, оптимальной длины линии $z_m$, которая и при $k=10$ оказывается очень большой. Вероятно именно это последнее обстоятельство является основным конструктивным ограничением максимально допустимой величины $k$.

## Литература